\documentclass[11pt,twoside]{article}
\usepackage[pdftex]{graphicx}
\usepackage{amsmath}
\usepackage{amssymb}
\usepackage{cite}
\usepackage{euscript}

 \newcommand{\tr}{\textup{Tr}}
 \newcommand{\hr}{\hat{\rho}}

\begin{document}
\newcommand{\pst}{\hspace*{1.5em}}

\newcommand{\rigmark}{\em Journal of Russian Laser Research}
\newcommand{\lemark}{\em Volume 30, Number 5, 2009}

\newcommand{\be}{\begin{equation}}
\newcommand{\ee}{\end{equation}}
\newcommand{\bm}{\boldmath}
\newcommand{\ds}{\displaystyle}
\newcommand{\bea}{\begin{eqnarray}}
\newcommand{\eea}{\end{eqnarray}}
\newcommand{\ba}{\begin{array}}
\newcommand{\ea}{\end{array}}
\newcommand{\arcsinh}{\mathop{\rm arcsinh}\nolimits}
\newcommand{\arctanh}{\mathop{\rm arctanh}\nolimits}
\newcommand{\bc}{\begin{center}}
\newcommand{\ec}{\end{center}}

\thispagestyle{plain}

\label{sh}


\begin{center} {\Large \bf
\begin{tabular}{c}
Tomographic Discord and Quantum 
\\[-1mm]
Correlations for System of Qubits
\end{tabular}
 } \end{center}

\bigskip

\bigskip

\begin{center} {\bf
V.I.Manko$^{1*}$ and Anatoli Yurkevich$^2$
}\end{center}

\medskip

\begin{center}
{\it
$^1$P. N. Lebedev Physical Institute, Russian Academy of Sciences,
Leninskii Pr. 53, Moscow, Russia 119991
\smallskip

$^2$Moscow Institute of Physics and Technology, Institutskii per. 9, Dolgoprudny, Moscow Region, Russia 141700
}
\smallskip

$^*$Corresponding author e-mail:~~~manko@sci.lebedev.ru \ anatoli.yurkevich@gmail.com
\\
\end{center}

\begin{abstract}\noindent
The difference of quantum mutual information for bipartite system of qubits and minimum taken with respect to local unitary transformation group is introduced as a characteristic of quantum correlations.The two qubits example (and specifically the two qubits X state)  is studied in some details.
\end{abstract}

\medskip


\section{Introduction}
\pst
 The classical and quantum systems containing subsystems have different properties,in particular, with respect to correlations of random observables. Such phenomenon as violation Bell inequalities \cite{1} and Tsirelson bound for system of two qubits is well known example of such difference.Also quantum discord \cite{2,3} is another example of the difference in properties of classical and quantum correlations.The entropic inequalities \cite{4} which recently were checked experimentally \cite{5} provide informational characteristics of the quantum correlations in the case of continuous variables. For the discussing the classical and quantum correlation the tomographic probability representation of states both in classical and quantum domains provides convenient tools \cite{6,7}.
  In this representation the quantum states are identified with fair probability distributions called tomographic probability distributions or tomograms. The tomograms can be introduced as notion of state also for classical systems \cite{8,6}.All the difference in behaviour of classical and quantum states is coded by the specific properties of the tomographic probability distributions.These properties are connected with reconstruction formulas for density operator which can be used not inly in quantum but also in classical mechanics \cite{9,6}. In quantum domain the density operator must be nonnegative and in classical domain it can have negative eigenvalues. This is the main reason that the classical and quantum correlations in component systems are essentially different. The aim of this work is is to study an analog of quantum discord which can be introduced for qubit systems on the base of unitary tomographic description of spin systems \cite{10,11}. The paper is organised as follows. In Sec.2 we describe the tomographic representation for spin states itself. In Sec.3
  we define the tomographic discord using the tomographic probability distribution and in Sec. 4 we apply it to two-qubit X-states.  
  
 \section{Tomographic Representation for Spin States}
 \pst
  Quantum state of the spin $ j=0,1\!/2,1,\dots $ with $m=-j,-j+1,\dots,j$ can be described with its tomogram $\omega(m,\textbf{n})$
 $$ \omega(m,\textbf{n})=\langle\delta(m-\textbf{n}\cdot\hat{\textbf{J}})\rangle=\tr{(\hr\delta(m-\textbf{n}\cdot\hat{\textbf{J}}))}, \eqno{(1)}$$
 where the vector $\textbf{n}=(\sin\theta\cos\phi,\sin\theta\sin\phi,\cos\theta) $ specifies a point on a sphere, $\hat{\textbf{J}}$ is the spin operator and where the Kronecker delta function 
 $$\delta(x)=\frac{1}{2\pi}\int\limits_0^{2\pi}e^{ix\phi}\,d\phi \eqno {(2)} $$
 is averaged by using the density operator $ \hat{\rho} $. The tomogram means the probability that the spin projection onto the direction of the vector $\textbf{n} $ is m. The inverse transformation is obtained in \cite{12} in matrix form. For the bipartite system of two spins with spins $j$ , the tomogram is defined as the joint distribution of the spin projections $ m^{(k)}$ onto the directions $\textbf{n}^{(k)}$ where $ k=1,2 $. The tomogram for a such state is given by 
 $$ \omega\left(m^{(1)},m^{(2)},\textbf{n}^{(1)},\textbf{n}^{(2)}\right)=\left\langle\prod_{k=1}^{N=2}\delta(m^{(k)}-\textbf{n}^{(k)}\cdot\hat{\textbf{J}}^{(k)})\right\rangle . \eqno{(3)} $$
 Now we can define a unitary spin tomogram for the spin $j$ as
 $$ \omega(m,u)=\langle m\vert u^+\rho u\vert m\rangle , \eqno{(4)}$$
 where $\rho$ is the density matrix of the state, $u$ is a unitary $ (2j+1)\times (2j+1) $ matrix and $m$ is the spin projection. If we choose the matrix $u$ as a matrix of an irreducible unitary representation of the rotation group, we will get Eq.(1). For the bipartite system we may define the analogous joint distribution
  $$\omega(m_1,m_2,u)=\langle m_1,m_2\vert u^+ \rho u\vert m_1,m_2\rangle . \eqno{(5)} $$
  As any density matrix $\rho $ of the size $k\times k$ may be represented in a form 
 $$\rho=u_0^+\textup{diag} \{\lambda_1,\lambda_2,\dots,\lambda_k\}u_0, \eqno{(6)}$$ 
 where $u_0$ is unitary matrix and columns of $ u_0^+ $ are normalized eigenvectors of $\rho$ to corresponding nonnegative eigenvalues $ \lambda_1,\lambda_2,\dots,\lambda_k\geq0 $ of $\rho$:
 $$ u_0^+=\begin{pmatrix}
             u_{11} & \dots & u_{1k} \\
             \vdots & \ddots & \vdots \\
             u_{k1} & \dots & u_{kk}
             \end{pmatrix};\quad
     \overrightarrow{u_0}^{+}(\lambda_1)=\begin{pmatrix}
             u_{11} \\
             \vdots  \\
             u_{k1} 
             \end{pmatrix},\dots,
       \overrightarrow{u_0}^{+}(\lambda_k)=\begin{pmatrix}
             u_{1k} \\
             \vdots  \\
             u_{kk} 
             \end{pmatrix}.  \eqno{(7)} 
  $$
 Thus, the tomogram may be represented in a form 
  $$ \omega(m_1,m_2,u)=\langle m_1,m_2|u^+u_0^+\textup{diag} \{\lambda_1,\lambda_2,\dots,\lambda_k\}u_0u|m_1,m_2\rangle, \eqno{(8a)} $$
  where $ k=(2j_1+1)\times(2j_2+1) $. Or 
  $$\overrightarrow{\omega}(\textbf{u})=|uu_0|^2\begin{pmatrix}
     \lambda_1\\
     \lambda_2\\
     \vdots\\
     \lambda_k
     \end{pmatrix} ,  \eqno{(8b)}
     $$   
     where $ |a|^2 $ for any matrix $a$ means
      $${|a|}^2_{jk}=|a_{jk}|^2. $$
   \section{Tomographic Discord}
   \pst
   Let`s consider a bipartite system described by the density matrix $\rho_{1,2}$. On one hand, we already have one definition of mutual information
   $$I(1,2)=S_1+S_2-S_{1,2}, \eqno{(9)} $$
   where $ S_{1,2}=-\textup{Tr}\rho_{1,2}\ln{\rho_{1,2}} $ is von Neumann entropy of the whole bipartite system and $ S_1=-\textup{Tr}\rho_{1}\ln{\rho_{1}}$, $ S_2=-\textup{Tr}\rho_{2}\ln{\rho_{2}}$ are von Neumann entropies of the corresponding subsystems which are described by partial traces $\rho_1$ and $\rho_2$ . But on the other hand, we can define mutual information using the tomographic probability distribution and the Shannon entropy \cite{13}. The subsystems we will describe respectively
   $$\omega(m_1,u)=\sum_{m_2}\omega(m_1,m_2,u) \qquad \omega(m_2,u)=\sum_{m_1}\omega(m_1,m_2,u). \eqno{(10)} $$
   And the second definition of mutual information is given by 
   $$ \EuScript I(u)= H_1(u)+H_2(u)-H_{1,2}(u), \eqno{(11)} $$
   where $H_{1,2}(u)$ is the Shannon entropy of the whole system and $H_1(u), H_2(u)$ are ones of the subsystems respectively:
   $$H_{1,2}(u)=-\sum_{m_1,m_2} \omega(m_1,m_2,u)\log \omega(m_1,m_2,u), \eqno{(12)} $$
   $$H_1(u)=-\sum_{m_1} \omega(m_1,u)\log \omega(m_1,u) \qquad H_2(u)=-\sum_{m_2} \omega(m_2,u)\log \omega(m_2,u) . \eqno{(13)} $$
   As it was shown in \cite{4} , the minimum value for the tomographic Shannon entropy $H(u)$ is von Neumann entropy $S$ and it is achieved when $u=u_0^+$ what is derived from Eq.(8b). Also, one can show that in the case of product unitary matrix $u=u_1 \otimes u_2, \quad \omega(m_1,u)=\langle m_1\vert u_1^+\rho_1 u_1\vert m_1\rangle$ where matrices $u_1$ and $u_2$ have the same dimensions as $\rho_1$ and $\rho_2$ respectively. It means an independent unitary transformation of the first subsystem neglecting the dependence on the second one. The same is true for the second subsystem. Also it means that the minimum values for $H_1(u)$ and $H_2(u)$ in case of product unitary matrix $u=u_1\otimes u_2$ are $S_1$ and $S_2$ respectively and are achieved with unitary matrices $u_1=u_{10}^+$ and $u_2=u_{20}^+$ diagonalizing the partial traces $\rho_1$ and $\rho_2$. In case of such diagonalizing matrices Eq.(11) reduces to :
   $$ \EuScript I(u_{10}^+\otimes u_{20}^+)= S_1+S_2-H_{1,2}(u_{10}^+\otimes u_{20}^+) \eqno{(14)} $$
   So we can define tomographic discord as the difference between two definitions of mutual information Eq.(9) and Eq.(11):
   $$ \EuScript D =I(1,2) -\EuScript I(u_{10}^+\otimes u_{20}^+)=H_{1,2}(u_{10}^+\otimes u_{20}^+)- S_{1,2}. \eqno{(15)} $$
   From the definition of the tomographic discord it follows
    $$ \EuScript D\geq 0 \eqno{(16)}$$
   
   \section{Tomographic Discord for Two Qubits X States}
   \pst
   Let`s consider a two-qubit X state with following density matrix 
   $$\rho_{1,2}= \begin{pmatrix}
     \rho_{11} & 0 & 0 & \rho_{14} \\
     0 & \rho_{22} & \rho_{23} & 0 \\
     0 & \rho_{32} & \rho_{33} & 0 \\
     \rho_{41} & 0 & 0 & \rho_{44}
                  \end{pmatrix}.  \eqno{(17)}$$
   It has following eigenvalues 
   $$ \lambda_{1,2}=\frac{(\rho_{11}+\rho_{44})\pm\sqrt{(\rho_{11}-\rho_{44})^2+4\vert\rho_{14}\vert^2}}{2} \qquad \lambda_{3,4}=\frac{(\rho_{22}+\rho_{33})\pm\sqrt{(\rho_{22}-\rho_{33})^2+4\vert\rho_{23}\vert^2}}{2}. \eqno{(18)}$$
  Its partial traces $\rho_1$ and $\rho_2$ are
  $$ \rho_1=\begin{pmatrix}
          \rho_{11}+\rho_{22} & 0 \\
          0 & \rho_{33}+\rho_{44}   
               \end{pmatrix}   \qquad
       \rho_2= \begin{pmatrix}
          \rho_{11}+\rho_{33} & 0 \\
          0 & \rho_{22}+\rho_{44}   
               \end{pmatrix}     . \eqno{(19)} 
     $$          
   The unitary matrices $u_{10}^+$ and $u_{20}^+$ which diagonalize these partial traces are identity matrices. So the tomographic Shannon entropy $H_{1,2}(u_{10}^+\otimes u_{20}^+)$ is given by
   $$H_{1,2}(u_{10}^+\otimes u_{20}^+)=-\sum_{k=1}^{4}\rho_{kk}\log\rho_{kk} . \eqno{(20)} $$
   So the tomographic discord for two-qubit state is 
   $$ \EuScript D= -\sum_{k=1}^{4}\rho_{kk}\log\rho_{kk} + \sum_{k=1}^{4}\lambda_k\log\lambda_k \eqno{(21)}. $$
   \section{Conclusion}
     \pst 
     To conclude we point out the main result of the work. For generic X-state of two qubits we calculated the tomographic discord ( see Eq.(21)) expressed in terms of the matrix elements of the state density matrix. We proved that the tomographic discord for the X-state equals to zero only in case of diagonal density matrix. In other cases it is positive and characterizes the quantum correlations. The connections with the quantum discord \cite{2} will be considered. The result can be compared with the experimental data \cite{3}.


\begin{thebibliography}{99}
 
  \bibitem{1} J.S. Bell, On the problem of hidden variables in quantum mechanics, Rev. Mod. Phys. \textbf{38}, 447 (1966)
   \bibitem{2} H. Ollivier and W.H. Zurek, Phys.Rev.Lett. \textbf{88}, 017901 (2002)
   \bibitem{3} E.I. Kuznetsova, A.I. Zenchuk, arXiv:1109.6420v1 [quant-ph] (2011)
   \bibitem{4} M.A. Man'ko, V.I. Man'ko,R.V. Mendes, J. Russ. Laser Res., \textbf{27}, 507 (2006)
   \bibitem{5} M. Bellini, A.S. Coelho, S.N. Filippov, V.I. Man'ko, A.Zavatta [quant-ph] arXiv:1203.2974 
   \bibitem{6} A. Ibort, V.I. Man'kob, G. Marmoc, A. Simonic, F. Ventrigliac arXiv:0904.4439 [quant-ph] (2009)
   \bibitem{7} S. Mancini, V.I. Man'ko1, P.Tombesi arXiv:quant-ph/9603002v1 (1996)
   \bibitem{8} O. Man'ko, V.I. Man'ko J. Russ. Laser Res., \textbf{18}, 407 (1997)
   \bibitem{9} O. Man'ko, V.I. Man'ko J. Russ. Laser Res., \textbf{25}, 477 (2004)
   \bibitem{10} V.I. Man'ko, G. Marmo, E.C.G. Sudarshan, F. Zaccaria arXiv:quant-ph/0310022 (2003)
   \bibitem{11} S.N. Filippov, V.I. Man'ko J. Russ. Laser Res., \textbf{30}, 129 (2009)
   \bibitem{12} V. I. Man'ko and O. V. Man'ko, JETP \textbf{85}, 430
(1997)
   \bibitem{13} C.E. Shannon, Bell System Techn. J.,\textbf{27}, 379, 623 (1948)
   \end{thebibliography}
\end{document}